# Single-branch Er:fiber frequency comb for optical synthesis at the $10^{-18}$ level

Holly Leopardi[1,2], Josue Davila-Rodriguez[2], Franklyn Quinlan[2], Judith Olson[1,2], Scott Diddams[1,2], and Tara Fortier[2]

[1]Department of Physics, University of Colorado Boulder, 440 UCB, Boulder, Colorado 80309, USA

[2]Time and Frequency Division, National Institute of Standards and Technology, Boulder, Colorado 80305, USA

email: holly.leopardi@nist.gov and tara.fortier@nist.gov

**Abstract**: Laser frequency combs based on erbium-doped fiber mode-locked lasers have shown great potential for compact, robust and efficient optical clock comparisons. However, to simultaneously compare multiple optical clock species, fiber laser frequency combs typically require multiple amplifiers and fiber optic paths that reduce the achievable frequency stability near 1 part in $10^{16}$ at 1s. In this paper we describe an erbium-fiber laser frequency comb that overcomes these conventional challenges and supports optical frequency synthesis at the millihertz level, or fractionally $3 \times 10^{-18} \tau^{-1/2}$, by ensuring that all critical fiber paths are within the servo-controlled feedback loop. We demonstrate the application of this frequency comb as a synthesizer for optical clocks operating across a wavelength range from 650 nm to 2100 nm.

1. INTRODUCTION

Optical atomic clocks, which provide both stable and accurate timing with up to 18 digits of resolution [1, 2], represent extremely sensitive tools with which to study fundamental physics. For instance, fine changes in the relative frequency of optical atomic clock transitions can be used to detect possible time variations of fundamental constants [3-7], changes in the gravitational potential of the geoid at the cm-scale [8], and have potential applications in dark matter and gravitational wave detection [9, 10]. However, these measurements often require the comparison of clocks based on different atomic species, whose transition frequencies can be separated by 100's of THz.

Optical frequency combs (OFCs), which act like optical frequency rulers, provide a means to bridge the gap in frequency between atomic clocks, while enabling both the relative comparison of optical atomic clock transitions, as well as absolute comparison to the current microwave atomic reference in $^{133}$Cs [11-13]. Although OFCs have been an enabling technology in optical clock development, their ability to synthesize optical and microwave frequencies from atomic frequency standards and references has led to a host of additional applications [14] including, but not limited to, atmospheric trace gas detection, calibration of astronomical spectrographs [15, 16], ultra-low-noise microwave generation [17, 18], optical time and frequency transfer [19-24], as well as microwave timing synchronization in large-scale science facilities [25-27].

In the past decade there have been significant improvements in the performance of optical reference cavities that serve as the local oscillator in optical atomic clocks. The development of high mechanical Q mirror coatings [28] and the operation of optical reference cavities at cryogenic temperatures [29] are projected to increase the stability of atomic clocks by nearly an order of magnitude over the current state-of-the-art. Consequently, high fidelity frequency synthesis with these optical frequency references will require optical frequency comb sources with equally good, or better performance.

Er:fiber-based optical frequency combs are widely used as optical synthesizers and dividers because they facilitate the possibility of compact design, as well as robust and turnkey operation [30, 31]. These advantages are in contrast to traditional Ti:sapphire-based OFCs, which have been the "gold-standard" in terms of performance [32-34], but do not offer the immediate possibility for robust or continuous long-term operation. One drawback of Er:fiber-based OFCs is their natively low average output power (< 100 mW) and repetition rate (< 250 MHz), that often requires them to employ separate amplifiers and nonlinear fibers to enable access to multiple optical frequency references [35, 36]. This is often required because it is difficult to optimize the broadened spectrum generated in highly nonlinear fiber (HNLF) for simultaneous detection of multiple optical beat signals. Although more convenient for signal optimization, this "multi-branch" configuration results in uncompensated fiber paths that are outside the feedback loop of the laser and hence limit the measurement stability of these OFCs near or above one part in $10^{16}$ at 1 second averaging [35, 37]. This level of residual noise is barely sufficient to support the current state-of-the art in optical atomic clocks [1, 2, 38, 39], and is above the level required to support the reported stability of cryogenic optical cavities [29]. A transfer oscillator scheme [40, 41] has been used to remove the noise of the uncompensated fiber paths but adds significant electronic and optical complexity to the system.

    Alternatively, to address the noise limitations of the multi-branch approach we have developed an Er:fiber based OFC that uses a "single branch" detection scheme, as shown in Figure 1. This approach uses a single optical amplifier and nonlinear fiber that are within the laser feedback loop and common to all optical paths. Consequently, this scheme enables the comparison of optical atomic frequency references with a 1 mHz frequency resolution in one second of averaging.

## 2. SINGLE BRANCH OPTICAL FREQUENCY COMB SETUP

The optical frequency comb (OFC) in our measurements is based on a self-referenced 180 MHz repetition rate Er:fiber ring laser whose pulse train is generated via nonlinear polarization evolution (NPE) [42]. The laser operates with an average output power of 90 mW and an optical bandwidth of ~80 nm FWHM directly from the laser cavity. In our experimental setup, the output of the laser is amplified in a polarization maintaining (PM) erbium doped fiber optical amplifier (EDFA), which was constructed with short fiber lengths, approximately 10 cm of PM single mode fiber at 1550 nm on the input and output of the amplifier, and with only 70 cm of PM erbium gain fiber. Pumped with two 980 nm laser diodes at approximately 700 mW each, the optical pulses at the output of the EDFA had an autocorrelation width of <70 fs and an average power of 250 mW. Light from the EDFA is then directly fiber coupled to 40 cm of PM-HNLF [43] with a total throughput of > 80%. Due to the relatively long length of HNLF, we still observe significant structure on the broadened output spectrum. However, the use of PM components enables stability of the power and spectrum shape output from the HNLF. The broadened octave-spanning spectrum, shown in Figure 2 a), produced from the single branch OFC spans from 980 nm – 2200 nm and enables simultaneous beat-note detection of $f_0$ as well as 6 optical frequency references; Ca atomic clock at 657 nm (from doubled 1314 nm), Yb optical lattice clock at 1157 nm, Al ion clock at 1070 nm, Hg ion clock at 1126 nm, Sr optical lattice clock at 1064 and the $Eu^{3+}$:$Y_2SiO_5$ spectral hole optical reference at 1157 nm. As seen in Figure 2 b), all the beat signals aside from that at 657 nm have greater than 30 dB SNR when measured with a 300 kHz resolution bandwidth, which allows for direct locking and counting of the optical beat signals without the need for tracking oscillators.

For stabilization of the OFC, and for the measurement and comparison of the different optical atomic clock lasers, the output spectrum from the PM-HNLF is split into separate optical interferometers to obtain a heterodyne beat signal against the different clock lasers (see Figure 1). To minimize the timing errors contributed by the optical interferometer paths, the entire OFC setup is enclosed in an acrylic box and the light output from the HNLF is launched into free space before being spectrally split among the different heterodyne interferometers. The maximum non-common optical interferometer path was less than 1 m. Separate PPLN crystals are used for doubling light at 1314 nm to access the Ca clock laser at 657 nm and for doubling 2100 nm for self-referenced detection of $f_0$. Dividing the spectrum at the output of the HNLF with dichroic beamsplitters and reusing light from the *f-2f* interferometer allows for efficient use of the broadened spectrum.

Stabilizing the OFC to an optical frequency reference requires locking both the laser repetition rate and the laser offset frequency. To stabilize $f_0$, the optical beat signal between doubled light at the low frequency end of the spectrum (2100 nm) is compared against that at the high frequency end of the spectrum at 1050 nm using the $f_0$ interferometer as depicted in Figure 1 b). Photodetection of the optical $f_0$ beat results in an RF signal that is filtered, amplified and compared against a hydrogen-maser referenced synthesizer. The resulting error signal is used in a feedback loop that actuates on a single OFC pump laser at 980 nm to control the laser gain with approximately 300 kHz of bandwidth. Similarly, the laser repetition rate is stabilized by photodetecting the optical heterodyne beat signal between one mode of the OFC and an optical frequency reference. The measured RF signal is filtered, amplified and compared against a second hydrogen-maser referenced synthesizer. The resulting error signal is sent to an intra-cavity EOM for fast control (~300 kHz bandwidth) of the laser cavity length and to a PZT to compensate long-range drift.

3. RESIDUAL COMB INSTABILITY MEASUREMENT

In the single branch measurement scheme for our Er:fiber based OFC, the amplifier and nonlinear fiber are common to all optical paths, and stabilization of $f_0$ and $f_{rep}$ compensates the noise of those optical components. However, any optical paths and electronics that are not common to the measurement of a second or third optical standard, as well as any imperfection in the stabilization of $f_0$ and $f_{rep}$ will result in frequency fluctuations that will contribute adversely to the measurement stability.

From the comb equation, the characteristic RF frequencies, $f_0$ and $f_{rep}$, as well as the optical heterodyne beat between a single mode, $N$, of the OFC and an optical frequency reference, $v_{meas}$, absolutely determine the frequency of the optical reference as follows:

$$v_{meas} = N \times f_{rep} \pm f_0 \pm f_{meas} \quad (1).$$

Imperfect stabilization of $f_0$ and $f_{rep}$ as well as the noise due to non-common mode paths will manifest themselves as fluctuations on $f_{meas}$ as follows:

$$<\delta f_{meas}^2> = <\delta f_0^2> + <\delta f_{lock}^2> + <\delta \varepsilon^2> + <\delta f_{\Delta L}^2> \quad (2).$$

Here, instability on the in-loop locked beat signal, $<\delta f_{lock}^2>$, measures imperfections in $N \times f_{rep}$, $<\delta f_0^2>$ measures imperfection in the feedback loop for stabilization of $f_0$, $<\delta \varepsilon^2>$ measures the total fluctuations added by the non-common mode electronics (i.e., synthesizers and amplifier chains), and $<\delta f_{\Delta L}^2>$ measures noise due to non-common mode optical paths. (The assumption in Eq. (2) that the separate noise terms are uncorrelated, particularly $f_0$ and $f_{lock}$, is justified by the measurements reported below.)

Figure 1 a) is a block diagram representation of how we visualize and measure the different optical and electronic instability contributions from the OFC setup. Figure 3 shows their respective contributions to the measurement instability with a normalization frequency of 282 THz. We evaluate the added optical noise of the OFC by counting a series of optically derived RF signals on a $\Lambda$-type counter [44] with approximately 12-digits fixed frequency resolution. Because of the inherent dead time, the counter cannot preserve phase coherence of the measured signals and hence the observed time-dependence of the instabilities were all limited to $\tau^{-1/2}$. Normalized to a 282 THz carrier, the counter itself contributed a measurement instability of approximately $1 \times 10^{-19}$ level at 1s when measuring a 10 MHz signal. Because of its fixed frequency resolution, the contributed measurement instability will scale linearly with frequency (e.g., when counting a signal at 100 MHz, the measurement instability will be $1 \times 10^{-18}$ at 1s.) Although the counters used did not all provide the ideal 12 digits of frequency resolution, we maintained a measurement limit on RF signals at least a factor of 2 below the noise of all of our measured signals.

To quantify the instability due to imperfect stabilization of the OFC we measured the in-loop beat signals $f_0$ and $f_{lock}$, whose instabilities were measured to be $1.40 \times 10^{-18}$ and $1.26 \times 10^{-18}$ at 1s, respectively on a 282 THz carrier. Although the synthesizers used as references and the amplifier chains (used to boost the RF beat signal from -60 dBm to 10 dBm) are within the laser feedback loops they are not common to the measurement of a second optical standard. The same is true for any uncommon optical paths. As a result, any frequency fluctuations from these non-common mode optical and electronic paths will be written onto the laser spectrum and will contribute added noise to the measurement of an out-of-loop optical frequency reference.

The noise in the RF amplifier chain was measured by splitting an RF signal from a maser-referenced synthesizer and sending one signal to a counter and the second through an amplifier chain to a second counter. By removing the correlated noise from the synthesizer we find that a single RF amplifier chain contributes an upper limit to the fractional frequency instability of $2 \times 10^{-19}$ at 1s on 282 THz. The synthesizers used in the measurement were referenced to a hydrogen-maser signal with a measured 10 MHz fractional frequency instability of $9 \times 10^{-13}$ at 1s. From this, we calculated that the total noise contributed from the three synthesizers (added in quadrature), used as references for the laser and for measurement of the optically-derived RF signals, contributed a total instability at a level of $7.3 \times 10^{-19}$ at 1s. Instabilities resulting from Doppler shifts and changes in the index of refraction of air from uncommon optical paths in the free space interferometers was determined by locking the $f_0$ optical beat signal measured on one interferometer and measuring the same beat signal on a second interferometer. Approximately 40 cm of differential free space path length contributed an instability of approximately $2 \times 10^{-18}$ at 1s.

To determine whether the above sources accurately describe the noise of the OFC setup, we measured the contribution from any additional noise sources: differential noise in the HNLF, noise induced from non-orthogonality in the $f_0$ and $f_{rep}$ locks or possible deviations from the ideal description of a frequency comb. This was accomplished by comparing the extreme ends of the broadened spectrum against a single optical frequency reference to determine the performance of the optical synthesis with the OFC across an optical octave. To do this we phase-locked a mode, $v_{2N}$, to a 282 THz optical frequency reference at the short wavelength end of the optical spectrum and then measured the beat signal between the same reference at 282 THz and the frequency doubled light from an optical comb mode, $v_N$, at 141 THz, (see Figure 2 a). To remove any sensitivity from out-of-loop optical paths, both the locked and measured beat signals were detected on the same photodetector. We compared the heterodyne beat from the doubled signal to the calculated quadrature sum, $<\delta f_0^2> + <\delta f_{lock}^2> + <\delta \varepsilon^2>$ and found that they agreed to within 1 part in $10^{-19}$. From this comparison we determined that there were no significant additional instability contributions from the OFC. Thus, using Eqn. 2, we calculate that the total measurement instability of our single branch OFC to a second out-of-loop optical measurement is approximately $3 \times 10^{-18}$ at 1s on 282 THz. The total measurement instability of our system can be further reduced by shortening the uncompensated free-space optical interferometer path lengths, or conversely, by measuring multiple beat signals on the same detector or purging the acrylic enclosure.

In addition to the stability, we can place an upper limit on the frequency offsets contributed by the single branch OFC at $4.8 \times 10^{-19}$ on a 282 THz carrier. This upper bound is currently limited by temperature-dependent frequency shifts of the RF synthesizers that are used for referencing and for measurement of the OFC. While the individual contributions to the offsets requires further evaluation, temperature control of the RF synthesizers, or additionally, by synthesizing the RF reference frequencies with the OFC itself, should reduce the total measurement error.

In order to demonstrate that the single branch OFC can support the current best optical frequency references at NIST, we performed an optical comparison between the Yb lattice clock laser at 1157 nm and the Al$^+$ quantum logic clock laser at 1070 nm. Approximately 3 mW of optical power from each optical standard was transferred via actively noise-cancelled fibers [44] across approximately 30-m of fiber. As seen in Figure 4, the optical comparison yielded a 1s fractional frequency instability of 3.84 x 10$^{-16}$ (linear drift of removed), limited by the current stability of the Al-clock laser.

## 4. CONCLUSIONS

We have developed and characterized the performance of a single branch Er:fiber frequency comb for the synthesis and comparison of optical atomic clocks. In our single branch configuration, the Er:fiber comb utilized a single PM EDFA and PM HNLF that were both within the laser feedback loop and common to all optical interferometers for stabilization of the offset frequency, as well as for characterization of the 6 optical standards at NIST. Due to the single branch design of the system, we were able to measure an out-of-loop residual instability of 1.78 x 10$^{-18}$ at 1s that averaged down as $\tau^{-1/2}$ on a 282 THz carrier. As summarized in Figure 4, this instability is nearly two orders of magnitude below the reported 1s instability of the current state-of-the-art in optical atomic clocks [1, 2, 38], and is one order of magnitude lower than the projected stability for next generation optical atomic clocks using a clock laser stabilized to a cryogenic optical cavity [29]. Additionally, the observed instability is better than that of a multi-branch Er:fiber comb system that requires a transfer oscillator technique for noise cancellation to attain measurement instabilities below 1 x 10$^{-16}$.

Finally, these measurements demonstrate that a single branch Er:fiber comb can perform at the similar level to OFCs based on Ti:sapphire lasers [32, 34]. As a result, an OFC based on an Er:fiber laser is an excellent potential candidate for use in an optical timescale for future redefinition of the SI second to an optical transition, and one that can support both the state-of-the art references of today, as well as those of the future.

## 4. ACKNOWLEDGEMENTS

This research was supported by NIST and the DARPA PULSE program. H. L. would like to thank NDSEG Fellowship for financial support. We would like to thank M. Hirano from Sumitomo Electric Inc. for the HNLF, and A. Ludlow, R. Brown, R. Fox, M. Schioppo, N. Hinkley, W. Mcgrew, R. Fasano, D. Nicolodi, G. Milani and D. Hume for providing access to the various optical frequency references.

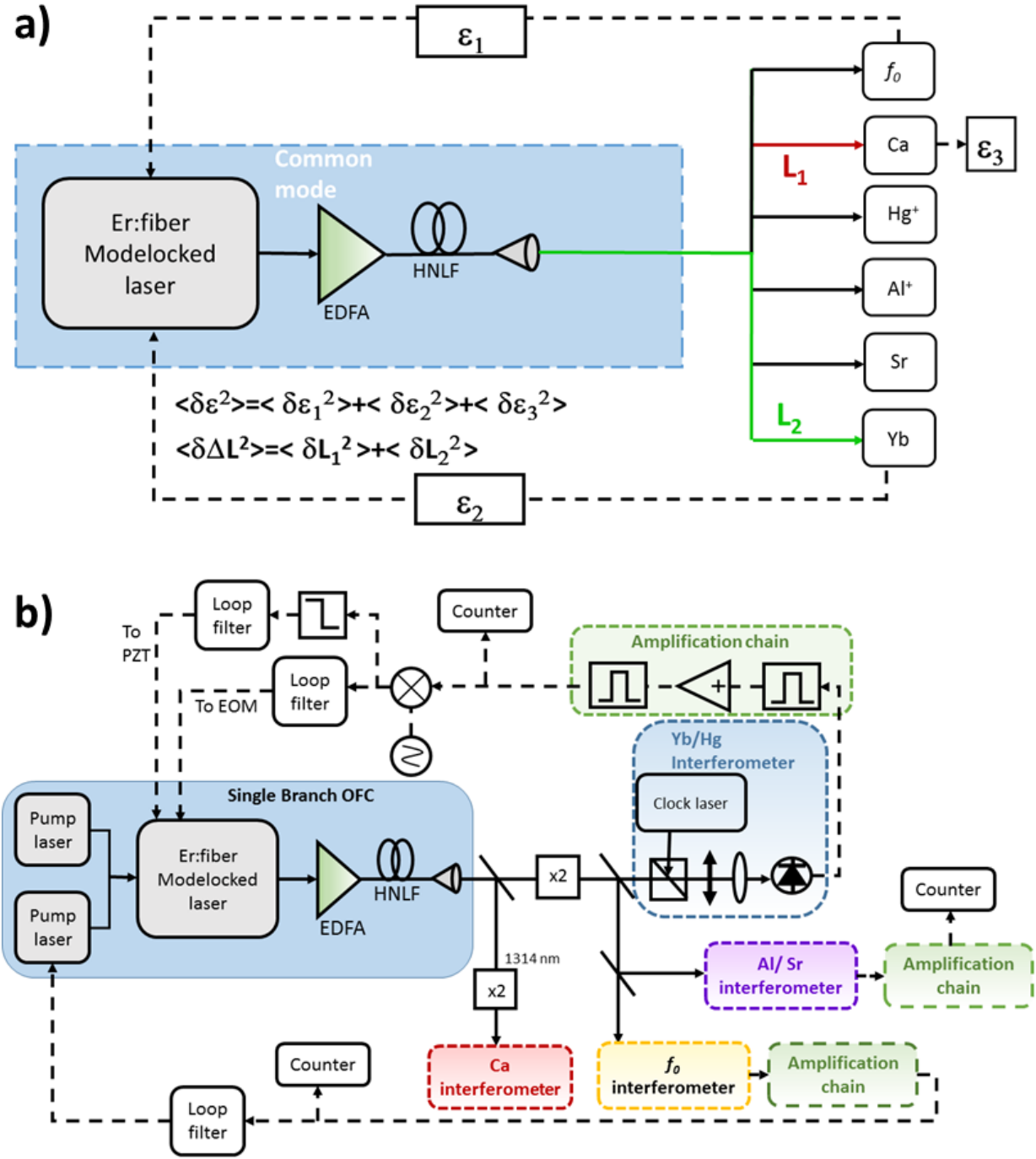

**Figure 1:** a) Block diagram highlighting the individual noise contributions within the optical frequency comb setup to measurement of an optical frequency reference at 282 THz. b) Simplified experimental setup for detection of the laser offset frequency and comparison to the 6 optical frequency references at NIST with a single branch Er:fiber-based frequency comb.

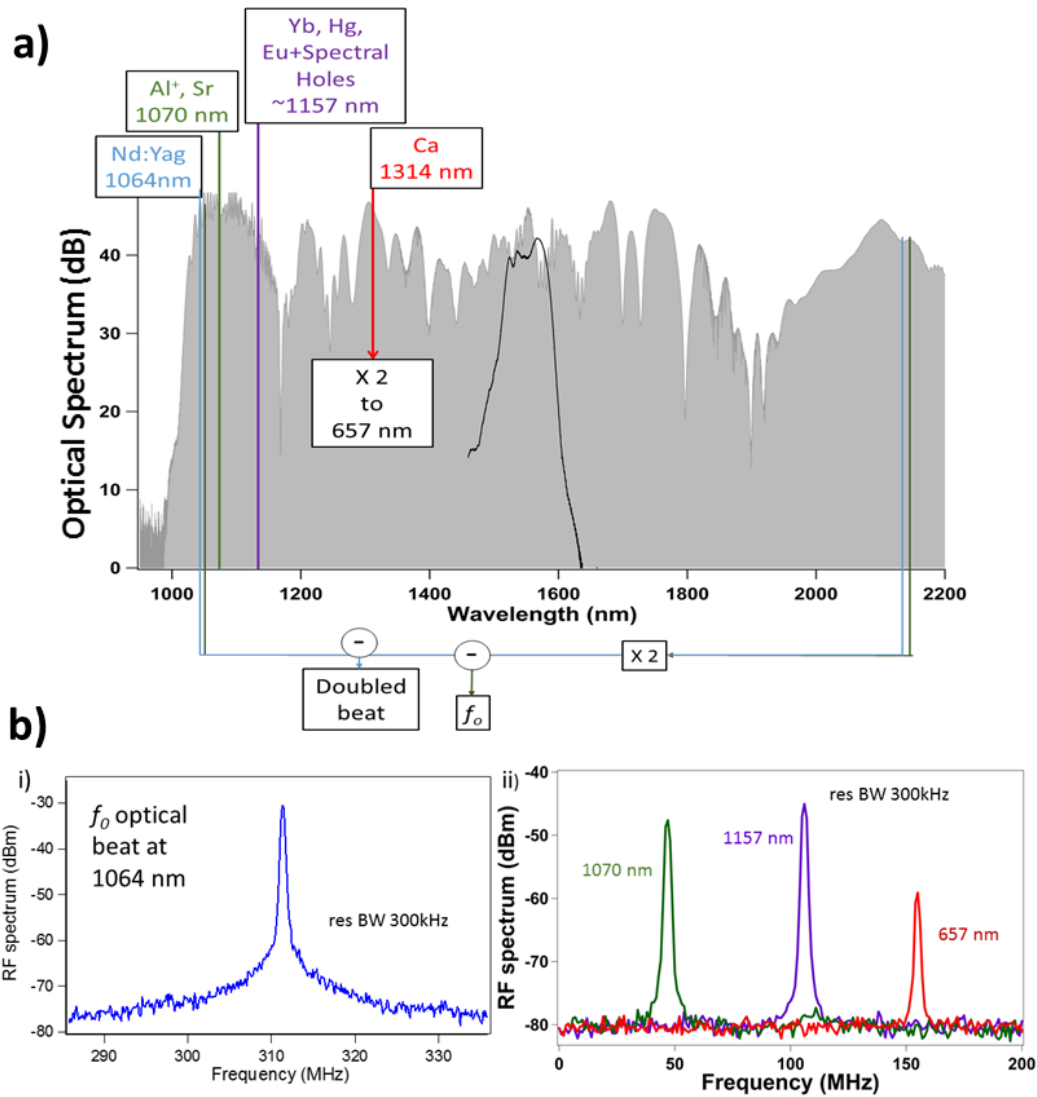

**Figure 2:** a) The optical spectrum output after optical amplification and external broadening in PM- highly nonlinear fiber and its overlap with the various optical frequency standards and references at NIST, Boulder. b) Optical heterodyne beat signals shown in a 300 kHz resolution bandwidth between the optical spectrum in Figure 2 and i) the offset frequency detected at 1050 nm and ii) the optical beats with the Al$^+$ clock laser at 1070 nm in green, the Ytterbium lattice clock laser at 1157 nm in purple, and Ca clock laser at 657 nm in red.

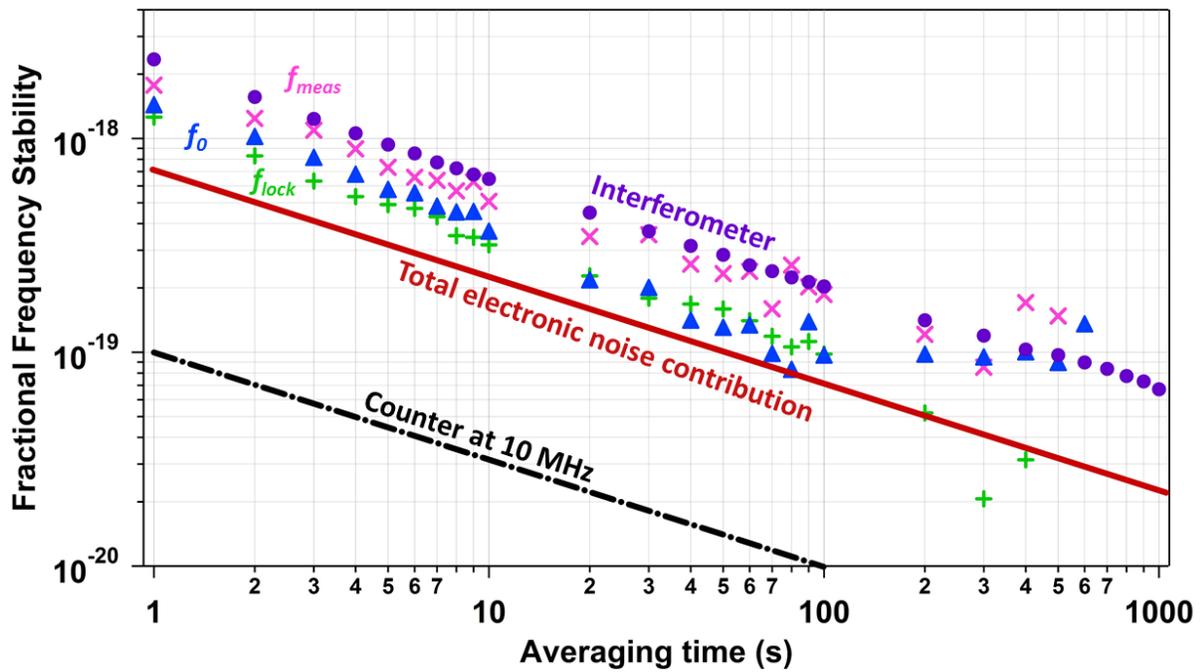

**Figure 3:** The itemized residual noise contributions of the single-branch OFC to an optical measurement at 282 THz. The blue triangles and green "+" measure the residual noise in the $f_0$ and $f_{lock}$ phase locked loops, respectively. The pink "x" measure the residual noise across the octave of an out-of-loop beat, $f_{meas}$. The purple solid circles shows the upper limit to the noise added by a 40 cm free space interferometer. The total contribution from amplifier chains and synthesizers is shown as the red solid line. The contribution of the counter instability measuring a 10 MHz signal is shown in black. The counter instability will scale linearly with the measured frequency.

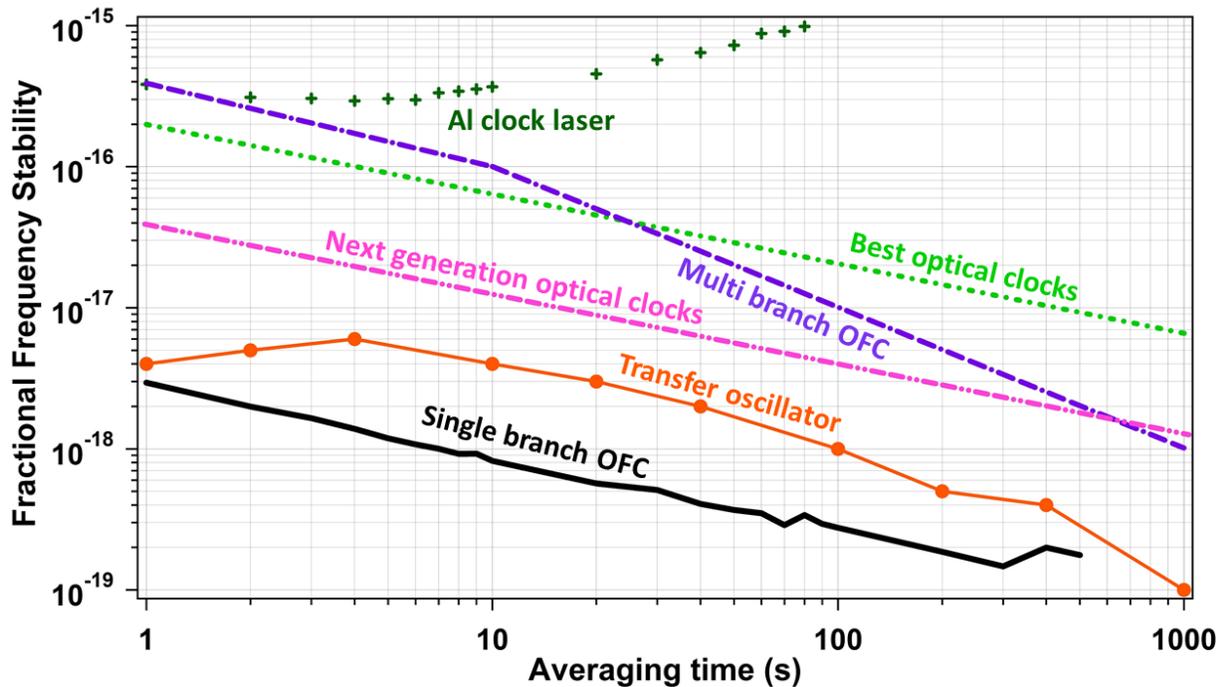

**Figure 4:** Residual instability of the single branch Er:fiber OFC (solid black line) compared to the instability of the multi branch Er:fiber OFC configuration (purple dashed line) [37, 45], the Er:fiber OFC with transfer oscillator technique applied (orange markers) [41], as well as the state-of-the-art optical cavities and optical atomic clocks (green dashed line) [1,2]. The pink dashed line represents the theoretical next generation optical atomic clocks based on a cryogenic optical cavity [29]. Additionally, we show the optical comparison between Yb lattice clock laser and Al$^+$ ion clock laser with the single-branch OFC (green crosses).